# Trouble of Non-Linearity

Lun-Shin Yao
Department of Mechanical and Aerospace Engineering
Arizona State University
Tempe, Arizona 85287

**Abstract**

All complex fluid motions, such as transition and turbulence, obeying the Navier-Stokes equations are *non-linear* phenomena. Some aspects of the non-linear terms of these equations are not well understood and are, in fact, misunderstood. The one-dimensional Kuramoto-Sivashinsky (KS) equation is used as a simple model non-linear partial differential equation to show some essential functions of its non-linear term and its consequences, which, we believe, are shared with other non-linear partial differential equations. We show that solutions of *nonlinear* partial differential equations above their critical parameters may be linearly stable, but are nonlinearly unstable. No stable solution exists above the critical parameter, contrary to the prediction of the linear-stability analysis. This is because a linearly stable disturbance can transfer energy *simultaneously*, not necessarily in cascade from small wave numbers to large wave numbers. An initial disturbance can breed its entire harmonics simultaneously. Second, we show that a long-time numerical chaotic solution cannot be achieved by a discrete numerical method.

## 1. Introduction.

We have shown that the nonlinear Navier-Stokes equations or other nonlinear partial differential equations (PDE) can have infinitely many equilibrium solutions; initial conditions determine which one is energized (Yao and Ghosh Moulic 1995a, Yao and Ghosh Moulic 1995b). We used the one-dimensional Kuramoto-Sivashinsky (KS) equation as a model to demonstrate that the nonlinear terms cause forced and resonant vibrations (Yao 1999). The forced vibration does not cause much energy transfer, and will not change the configuration of the solution, but excites all harmonics. A substantial amount of energy is transferred due to resonance and can completely alter the configuration of the solution; consequently, the non-linear terms cannot be ignored even if their amplitudes are small. The resonance conditions are usually satisfied for commensurate waves; consequently, they are self-similar, but not topologically transitive. This means that the solution is sensitive to the initial conditions and can have many different long-time solutions due to different initial conditions.

Since all complex flow phenomena are related to the non-linear terms of the Navier-Stokes equations and the role of these terms is often misunderstood, we use the KS equation as a simple model partial differential equation to clarify the role of non-linear terms and their consequences. We do not intend to study the solution properties solely for the KS equation, which have been intensively studied, and are considered well known (Jolly, Kevrekidis and Titi 1990). We believe that the numerical examples discussed below demonstrate general properties, not just for the KS equation, but for all non-linear partial differential equations including the Navier-Stokes equations.

First we want to show that a linearly stable or unstable initial condition always excites its harmonics, no matter how small their amplitudes are. Therefore, nonlinear PDE have no stable long-time solution if the parameter is larger then its critical value. At the critical point, the influence of the linear term vanishes for the critical wave, but it can transfer energy non-linearly and simultaneously, not in cascade, to its harmonics which are dissipative. This causes the amplitude of the critical wave to approach zero asymptotically and is not neutrally stable as predicted by the linear-stability analysis. Consequently, a linear-instability boundary lacks physical significance in the presence of non-linear terms. Another natural consequence is that a *single-wave* solution does not exist for a non-linear partial differential equation. This also provides a simple explanation as to why a long wave can trigger short Tollmien-Schlichting waves in boundary-layer transition. Since energy is not necessarily transferred from long waves to short waves in cascade, short waves are directly influenced by the long waves and are unnecessary to be isotropic.

Second, we show that long-time numerical chaotic solution is impossible by a time-discrete method. This is due to the fact that the Liapunov exponent is positive for chaos. However small, truncation or round-off errors can be exponentially amplified by the non-linear evolution . Hence, numerical solutions depend on the integration time step, and cannot be interpreted as a solution of the differential equation. This implies that a direct numerical solution of the Navier-Stokes equations for turbulence is not possible until the next generation numerical methods can be developed, which are truncation-error free.

## 2. Analysis and Results

A spectral method is used to a spatially discrete form of the KS equation

$$u_t + 4u_{xxxx} + l\,\left(u_{xx} + uu_x\right) = 0, \tag{1}$$

for x in the interval [0,2π], where λ is the governing parameter and signifies the energy production and non-linear energy transfer. An approximate solution is written as

$$u(x,t) = \sum_{n=1}^{N}\left(A_n e^{inx} + A_{-n} e^{-inx}\right) , \tag{2}$$

where $A_{-n}$ is the complex conjugate of $A_n$. An small-system example for N = 5 is

$$\frac{\partial}{\partial t}\begin{bmatrix} A_1 \\ A_2 \\ A_3 \\ A_4 \\ A_5 \end{bmatrix} = \begin{bmatrix} 1^2\left(\lambda - 4 \bullet 1^2\right) & & & & \\ & 2^2\left(\lambda - 4 \bullet 2^2\right) & & & \\ & & \bullet & & \\ & & & \bullet & \\ & & & & 5^2\left(\lambda - 4 \bullet 5^2\right) \end{bmatrix}\begin{bmatrix} A_1 \\ A_2 \\ A_3 \\ A_4 \\ A_5 \end{bmatrix} - i\lambda \begin{bmatrix} 0 & A_{-1} & A_{-2} & A_{-3} & A_{-4} \\ A_1 & 0 & 2A_{-1} & 2A_{-2} & 2A_{-3} \\ A_2 & 2A_1 & 0 & 3A_{-1} & 3A_{-2} \\ A_3 & 2A_2 & 3A_1 & 0 & 4A_{-1} \\ A_4 & 2A_3 & 3A_2 & 4A_1 & 0 \end{bmatrix}\begin{bmatrix} A_1 \\ A_2 \\ A_3 \\ A_4 \\ A_5 \end{bmatrix} \quad \textbf{(3)}$$

which clearly shows that the energy transfers from small wave numbers to large wave numbers in sequence is one of many possible paths, and is not the only one; moreover, it is not necessarily the dominant one. It also shows that the only small amount of energy transfers non-linearly to large wave numbers since it is amplitude dependent; therefore, the amplitudes of large-wave numbers are small. Energy transfer is usually from high amplitude waves to low amplitude ones, but does not necessarily always follow this rule. Since the wave with n = 0 does not interact with others, it is not included in the computation. This is a limitation of the one-dimensional model. Otherwise, it is similar to the governing equation of the amplitude-density functions of a Fourier-eigenfunction spectral method for the Navier-Stokes equations (Ghosh Moulic and Yao 1996, Yao and Ghosh Moulic 1995a, 1995b). The linear terms of (3) represent the difference of energy production and dissipation. If its value is positive, then the mode grows after being excited and is known as linearly unstable.

Two time-integration schemes are used for the temporal discretization. The first method approximates the non-linear term by an explicit second-order-accurate Adams-Bashforth scheme and the linear term by an implicit Crank-Nicholson scheme and is referred to as the explicit method below. The implicit method is a second-order Crank-Nicholson scheme. A careful comparison of many numerical computations show that N=32 is more than sufficient for the two examples shown in the paper.

The first case is for $\lambda = 4$, where the wave $A_1$ is neutrally stable according to a linear-stability analysis. The initial condition is $A_1 = 1$ and the other waves are zero. The time step is 0.001. The energy cascade is clearly shown by the explicit method in Figure 1. The first time step $A_2$ starts to grow; the second time step, $A_3$ and $A_4$ are excited, and so forth. This is purely due to the limitation of the explicit method. It has been shown that the non-linear terms represent many wave resonances and are not just limited to resonant trios (Yao and Ghosh Moulic 1995b). The implicit method show that the energy transfers from $A_1$ to its harmonics *simultaneously* and is not cascaded from small wave numbers to large wave numbers. The energy transfer from $A_1$ causes the initial growth of the harmonics. All waves start to slowly declines after 60 time steps as shown in Figure 2. Consequently, the critical wave is not neutrally stable at the critical point if the non-linear effect is considered.

Since a long wave can simultaneously transfer energy to its entire set of harmonics, it becomes clear that short Tollmien-Schlichting waves triggered by a long wave are a natural non-linear process. This example also demonstrates that short waves can receive energy directly from long waves; hence they are unnecessarily to be isotropic.

The second case is for $\lambda = 69$, for which it has been shown that the solution is chaos (Jolly, Kevrekidis and Titi 1990). We found that an initial condition of a single wave will not cause a chaotic solution. If the initial amplitudes of linear unstable waves, n = 1- 4, are set to a very small number, say $10^{-10}$, then the numerical solution becomes chaotic. We also found that solutions are time-step sensitive after a very short time, say $t \geq 0.41$. The results presented below are for the initial amplitude of all waves at $10^{-5}$. Since the difference of production and dissipation of $A_3$ has a maximum value at $\lambda = 69$, we only plot the numerical result associated with $A_3$. The results of the other waves are similar, so that it is not necessarily to show them.

The time history of $A_3$ is plotted in Figure 3 for five different $\Delta t$'s by the explicit method. We found that the numerical result grows without bound for $\Delta t > 0.0007$. Reduction of the time step to $10^{-4}$ extends the convergent solution from t=0.2 to t=0.4. Further reduction of $\Delta t$ to a value smaller than $10^{-4}$ does not improve the results. The solution becomes $\Delta t$-sensitive after $t \simeq 0.41$. We have tried Adam-Bashforth methods up to 5th order, implicit method, fourth-order Runge-Kutta method as well as compact time-difference schemes (Yao 1998) and cannot obtain a convergent solution after $t \simeq 0.41$. The result is surprising, but not unexpected. Since a necessary property of chaos is a positive Liapunov exponent, or a positive non-linear exponential growth rate, the different truncation errors introduced by different $\Delta t$'s of various methods grow unstably at their own path. They can be viewed as the solution of the difference equation with selected $\Delta$t, but cannot be interpreted as a solution of the differential equation. A plot of time-averaged energy spectrum $E(n=3)=\frac{1}{t}\int_0^t \left|A_n^2\right| dt = \frac{1}{M}\sum_{m=1}^{M} A_n(m\bullet\Delta t)\bullet A_{-n}(m\bullet\Delta t)$ in Figure 4 shows that the time-averaged solutions also depend on the time-step. The case labeled FFT is by the psedospectral method. Since the Liapunov exponent of turbulence must be positive, it is doubtful that a direct numerical solution of turbulence is possible by the current generation of discrete numerical integration methods, which are not truncation-error free.

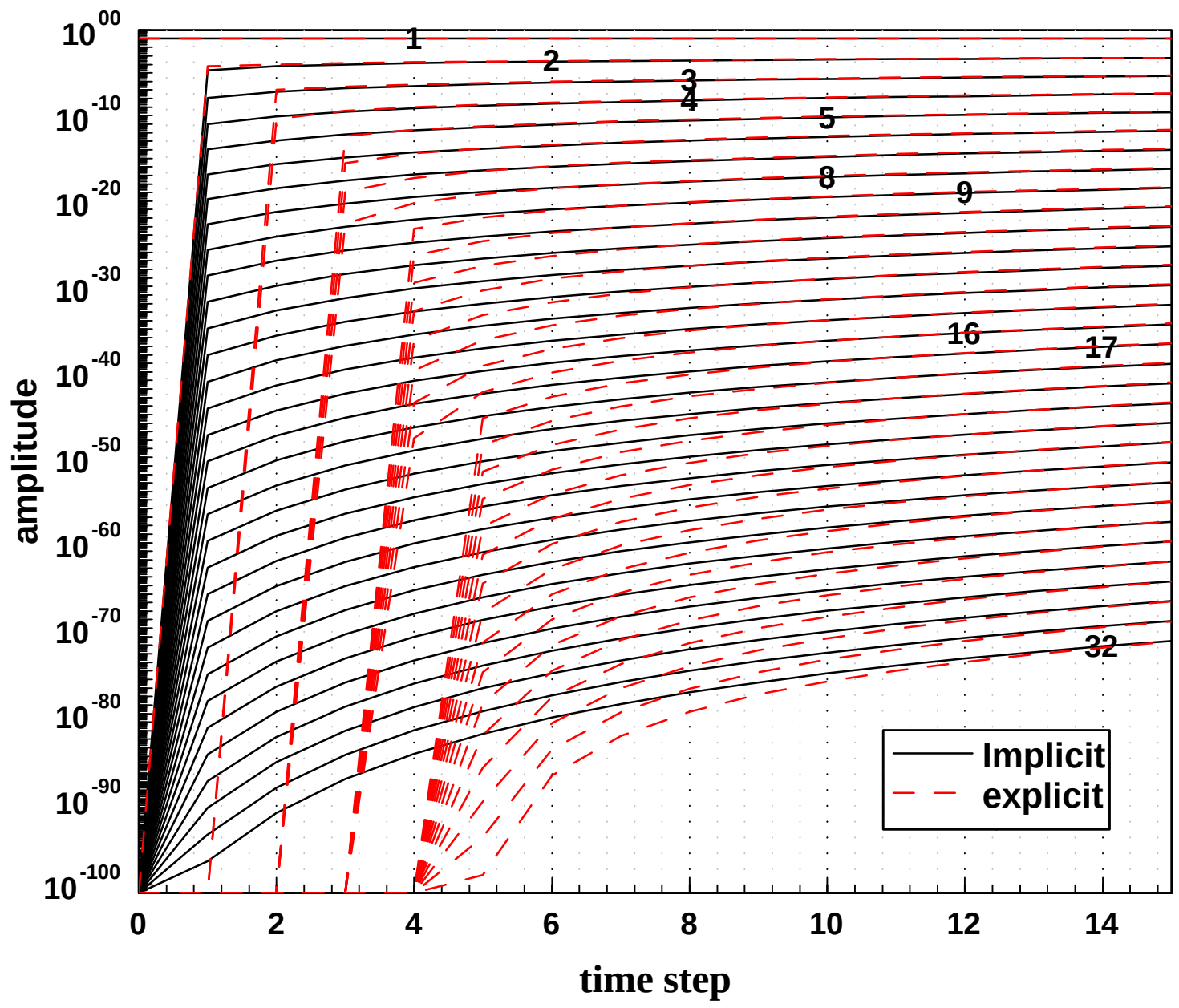

Figure 1. Non-linear evolution at the critical point for the initial condition that $A_1$=1.

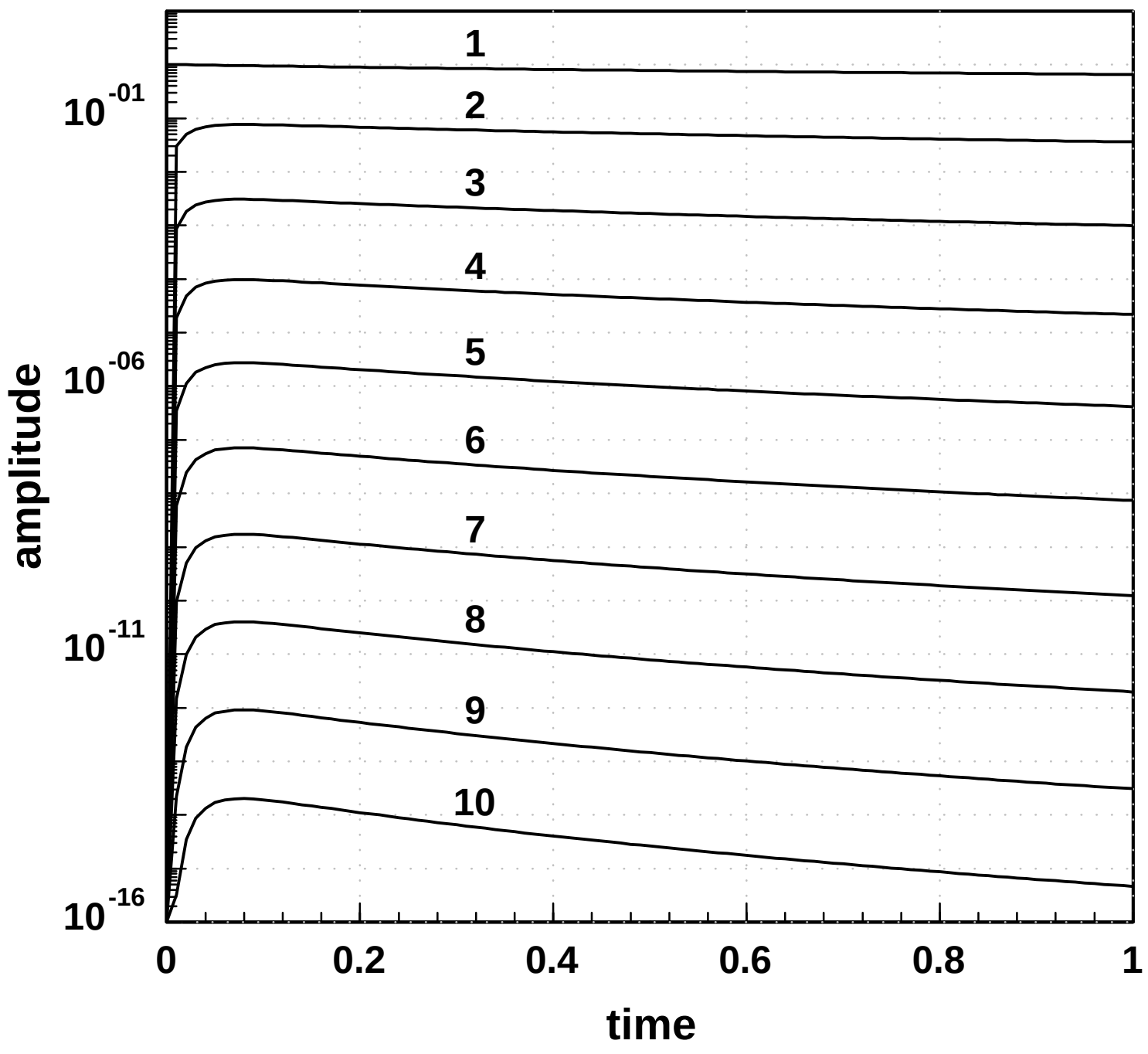

Figure 2. Non-linear evolution of the first ten waves at the critical point for the initial condition that $A_1$=1. $\Delta t$=0.001

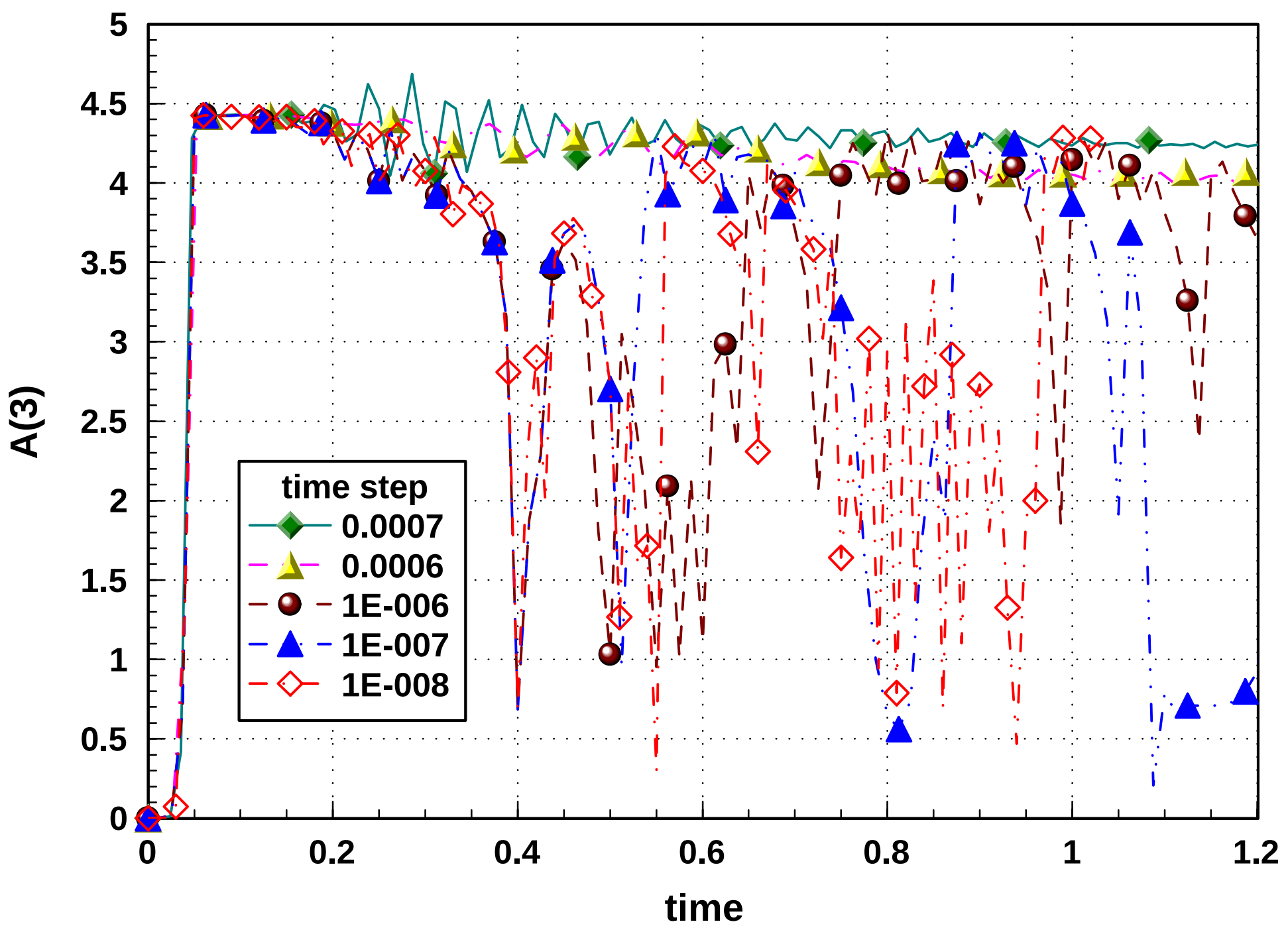

Figure 3. Time history of the amplitude of the third wave for various time steps

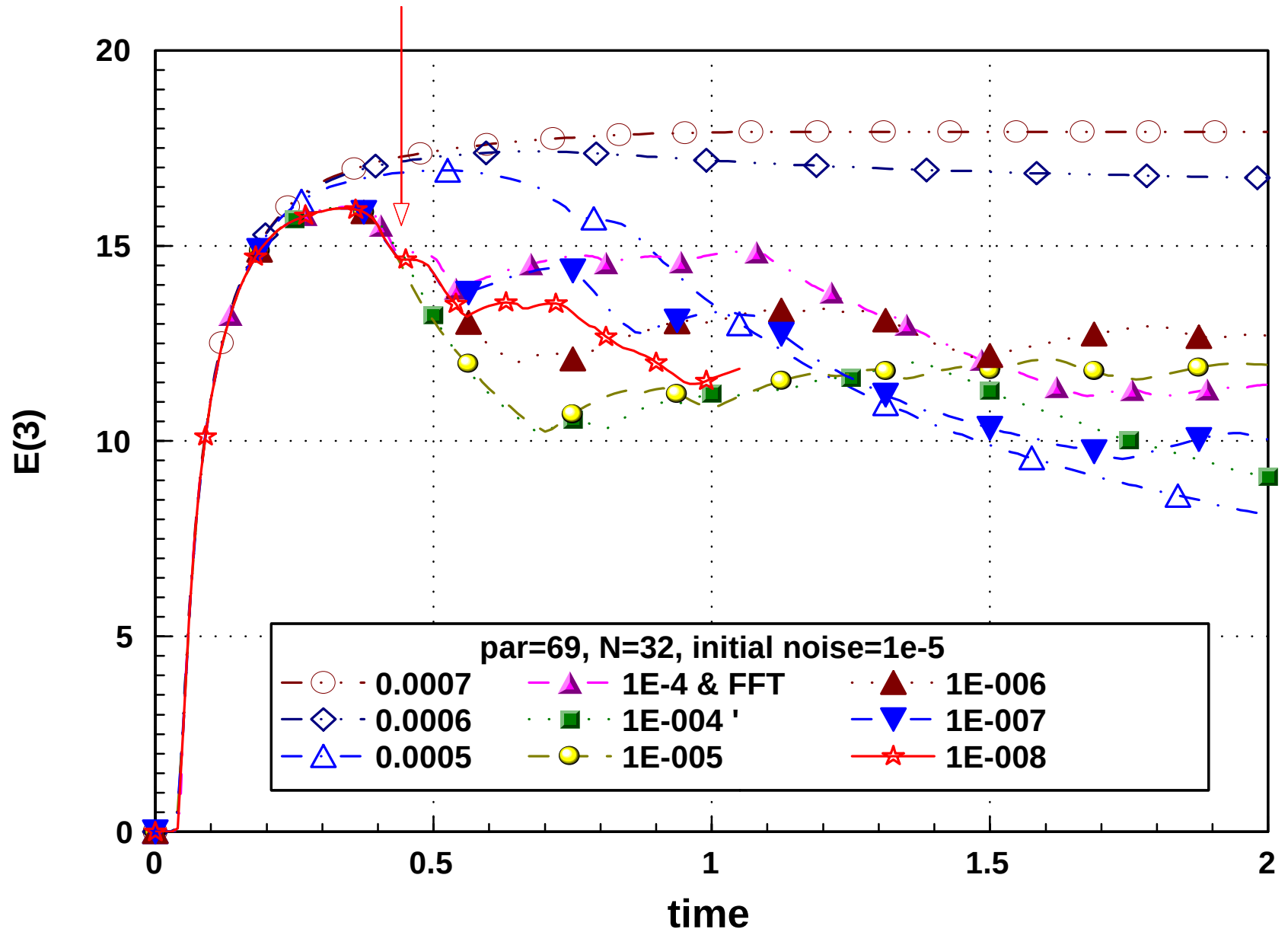

Figure 4. Time-averaged energy for the third wave for the various time steps

## Attachment: Communication with Professor Friedrich Busse, An Associate Editor for the Journal of Fluid Mechanics.

**From:** Lun-Shin Yao
**Sent:** Monday, January 28, 2002 2:49 PM
**To:** 'Friedrich Busse'
**Cc:** 'sdavis@casbah.it.northwestern.edu'; 'T.J.Pedley@damtp.cam.ac.uk'
**Subject:** RE: your mail

Dear Professor Busse:

I would like to re-emphasize again that **NUMERICAL SOLUTIONS FOR INSTABILITIES ARE POSSIBLE and DIRECT NUMERICAL SOLUTIONS FOR TURBULENCE ARE IMPOSSIBLE** with our current knowledge. All direct numerical solutions for turbulence, including **LARGE EDDY SIMULATIONS** are numerical errors. The **SPURIOUS** solutions associated with under-resolved computation are not the subject discussed in my paper.

Please read my two papers on numerical instabilities in which we have carefully checked the convergence of the solutions :

1. L.S. Yao and Sandipan Ghosh Moulic, "Nonlinear Instability of Traveling Waves With a Continuous Spectrum," *Int. J. Heat & Mass Transfer*, **38**, 1751-1772, 1995.
2. Sandipan Ghosh Moulic and L.S. Yao," Taylor-Couette instability of traveling waves with a continuous spectrum," *J. Fluid Mech.*, **324**, 181-198, 1996.

My 1996 paper is for **THREE-DIMENSIONAL** Taylor-Couette flows. We integrated the three-dimensional Navier-Stokes equations **IN TIME** to show that the initial condition determines the dominant equilibrium wave solution. I believe this is the first analytical/numerical paper to show that multiple equilibrium solutions exist for Taylor-Couette flows, which verifies Cole's 1965 experimental observation and provides a rational explanation for it.

In the above two papers, we used numerical results and perturbation analyses to demonstrate the inconsistency of the **WEAKLY NON-LINEAR THEORIES** and explained why a single-wave solution cannot exist for the non-linear Navier-Stokes equations, even though weakly non-linear theories have been very popular for more than forty years. Even today, such erroneous solutions are frequently published in JFM. The first half of my current paper, "Non-Linear Energy Transfer," used the simple K-S equation to re-demonstrate the reason why commonly accepted concepts are incorrect since my 1995 paper may be too complex for impatient readers to understand.

If you have any questions after you read my paper, please feel free to ask.

Sincerely,

Lun-Shin Yao

-----Original Message-----
From: Friedrich Busse [mailto:Friedrich.Busse@uni-bayreuth.de]
Sent: Saturday, January 26, 2002 8:57 AM
To: Lun-Shin Yao
Cc: 'Friedrich Busse'; 'T.J.Pedley@damtp.cam.ac.uk';
'sdavis@casbah.it.northwestern.edu'
Subject: RE: your mail

Dear Professor Yao,
I doubt that you ever have integrated in time a nonlinear system of 3d-partial
differential equations. But this is the task faced by fluid dynamicists.
Sincerely, F. Busse

On Fri, 25 Jan 2002, Lun-Shin Yao wrote:

> Dear Professor Busse:
>
> I have to say that I am continuously amazed by your comments.  I use
> computers to do my research, too.  All of my computation of flow
> instabilities are carefully checked and represent convergent solutions.  I
> will never try to publish erroneous results intentionally; in particular, I
> will not send my results out when I know there are bugs in my computer code.
> I also teach my students not do it. One exception is the chaotic solution of
> the Kuramoto-Sivashinsky equation in my 1999 JFM paper.  In the current
> paper, I tried to correct my earlier mistake.  Since I found that the
> numerical chaotic solutions of the Lorenz system are also numerical errors,
> there is little doubt in my mind that currently there is no way to construct
> chaotic solutions for non-linear differential equations.
>
> On the contrary, you are quite comfortable to take erroneous solutions as
> the proper solutions of your research problems.  It is clear that your
> research attitude is completely different from mine.  Because of this
> difference, my error becomes your solution.  Your attitude and limited
> knowledge allow you to create an impossible solution from numerical errors
> with confidence, and think this is the only way to practice CFD!  On the
> other hand, I am glad to see that you have learned something about
> attractors and CFD in our communication.  I am sure that JFM must be very
> proud to have an editor like you to maintain the journal standard.
>

> Sincerely,
>
> Lun-Shin Yao
>
> -----Original Message-----
> From: Friedrich Busse [mailto:Friedrich.Busse@uni-bayreuth.de]
> Sent: Thursday, January 24, 2002 3:15 AM
> To: Lun-Shin Yao
> Cc: 'Friedrich Busse'; 'T.J.Pedley@damtp.cam.ac.uk'
> Subject: RE: your mail
>
>
> Dear Professor Yao,
> I may not be the best editor in terms of Knowledge of CFD among my
> colleagues, but I have used electronic computers in most of my research
> since 1960. On the contrary, I get the impression from our discussion that
> you are not familiar with the practice of CFD. It is just not possible in
> practice "to show that the discretized numerical solution is
> asymptotically approaching the solution of a ( 3d-partial) differential
> equation by changing grid sizes and time steps systematically...."
> Since research is carried out at the limit of available computer capacity
> you will elicit only a tired smile from CFD-people when you confront them with
> your demands.
> Of course, there are probably quite a few results published in journals
> which should be corrected. With the increasing computer capacity this can
> be easily done in particular striking cases. Actually I think that bugs in
> computer codes are an even more serious problem than insufficient
> numerical convergence.
> You can appeal my decision by directly writing to Profs. Davis or Pedley.
> Yours sincerely, F. Busse
>
>
> On Wed, 23 Jan 2002, Lun-Shin Yao wrote:
>
> > Dear Professor Busse:
> >
> > I am bothered by your comments!  From the comments quoted below, it is
> > clear that you have changed your viewpoint about CFD:
> >
> > On 1/17/2002,
> > "Fluid dynamicists are well aware of the difficulties they face in
> > simulating
> > turbulent flows. Fortunately the chaotic attractors with which they deal
> > are quite robust such that simulations can describe the physics correctly
> > even after long times in spite of failing classical criteria for numerical
> > convergence."

> > On 1/20/2002,
> > "People in this area base their confidence in their
> > results at best on emperical experience. When changes in gridsize and
> > timestep do not cause any significant change in the results of the
> > computations, they will be satisfied."
> >
> > Surprisingly, you did not change your conclusion about my paper!
> >
> > I know you realize that we are not discussing politics or religions; on the
> > contrary, we are discussing SCIENTIFIC computation.  There is no room for
> > empirical experience.  In science, we should only trust EVIDENCE. Everyone
> > who does numerical computation knows that divergent numerical solutions
> > are numerical artifacts, not solutions to the governing differential equations.
> > One has to show that the discretized numerical solution is ASYMPTOTICALLY
> > approaching the solution of a differential equation by changing grid sizes
> > and time steps systematically in order to demonstrate the convergence of
> > his method for his problem.  Arbitrarily changing grid size and time step until
> > there is no significant change in the results of a computation is not a
> > satisfactory practice.   I have never heard the argument that individuals
> > can "base their confidence in their results at best on empirical
> > experience".  From your comments, it is clear that your expertise is not
> > in CFD.  Your belief would imply that an experimentalist can accept a
> > measurement as long as he has confidence in his results because of his
> > empirical experience, thus not having to calibrate his instruments.
> > Following your belief we would only need to make a mistake once, and will
> > never bother to find out if it is a mistake or not.  What a nice world it is!
> >
> > Your comment that a correct CFD approach must occur in journals of CFD or
> > numerical analysis also bothered me.  I don't think that you intend to imply
> > that erroneous numerical solutions are quite acceptable in JFM.  If anyone
> > wants to correct it, he can only publish his paper in journals of CFD, and
> > not in JFM.  I certainly hope that the editors and other associated
> > editors of JFM do not follow your logic.
> >
> > My paper addresses the numerical difficulty concerning the general mistake
> > made by all papers of "direct computational turbulence."   I do not see why
> > I should single out one of them by submitting a discussion comment.  Let me
> > be frank.  I have not seen any direct numerical turbulence paper published
> > in JFM that demonstrates its convergence.  I will greatly appreciate it if
> > you can point out one paper which presents a CONVERGENT direct numerical
> > turbulence.  I will be eagerly to read it.
> >
> > I do not know the policy of JFM very well, but I would like to suggest you
> > to pass my paper back to Tim Pedley, since I do not believe that CFD is
> > your research area.  Also, I do believe that Cambridge group should have
> > experts in this area, who can fairly judge my paper.

> >
> > Sincerely,
> >
> > Lun-shin Yao
> >
> > -----Original Message-----
> > From: Friedrich Busse [mailto:Friedrich.Busse@uni-bayreuth.de]
> > Sent: Sunday, January 20, 2002 2:59 AM
> > To: Lun-Shin Yao
> > Cc: 'Friedrich Busse'; 'T.J.Pedley@damtp.cam.ac.uk'
> > Subject: RE: your mail
> >
> >
> > Dear Professor Yao,
> > I sympathize with your dissatisfaction with the current state of the
> > foundations of CFD. People in this area base their confidence in their
> > results at best on emperical experience. When changes in gridsize and
> > timestep do not cause any significant change in the results of the
> > computations, they will be satisfied.
> > A better mathematical underpinning of CFD is desirable but that
> > development must occur in journals of CFD or numerical analysis. You may
> > also consider the option of writing a comment on a paper which, in your
> > opinion, contains particular glaring incorrect results and misleading
> > conclusions based thereupon. Those comments usually receive much more
> > attention. But I still think that in its present form your paper is not
> > appropriate for publication in JFM.
> > Yours sincerely, F. Busse
> >
> >
> > On Thu, 17 Jan 2002, Lun-Shin Yao wrote:
> >
> > > Dear Professor Busse:
> > >
> > > I am surprised to read your email.  You implied that your referee does not
> > > have to understand the content of my paper.  I want to point out another
> > > incorrect comment he or she made about my paper.  The results for the K-S
> > > equation in my 1999 JFM paper show the chaos is due to resonance, but,
> > > at that time, I did not recognize my mistake in failing to appreciate the
> > > significance of  numerical errors.  The current paper demonstrates that
> > > the computed time-averaged kinetic energy differs for different time steps
> > > and different finite-difference methods.  Please tell me which time step or
> > > method I should use?  Can I pick one which agrees with experimental
> > > observation and ignore others?  The kinetic energy is an important physical
> > > quantity in turbulence.  I did compute a long-time average up to t=500
> > > and did not find a convergent solution.  I hope that you realize that your
> > > comment about my paper having little to do with fluid dynamics is

> > > shortsighted.  Every issue discussed in my paper is related to fluid
> > > mechanics.
> > >
> > > I too am disappointed that finite-difference solutions contain non-trivial
> > > numerical errors.  This is because  our original goal was to attempt to
> > > demonstrate that there are multiple equilibrium turbulence structures
> > > for a given Reynolds number and geometry, as has been verified for
> > > supercritical instabilities, but not for turbulence; also, turbulence is not ergodic.
> > > Right now, we do not have a mathematical tool to achieve this goal, and
> > > experimental accuracy is not likely to come in the near future.
> > >
> > > Many researchers in CFD believe that numerical chaotic solutions are indeed
> > > approximate solutions of these nonlinear differential equations, if there is
> > > a true attractor.   I believed that too until I found the time-averaged L2
> > > norms (or turbulent kinetic energy) are different for different time steps.
> > > I believe that attractors provide visual pleasure, but time-averaged
> > > statistical quantities have more physical relevance.   I further believe
> > > that anyone who has experienced what I did has to conclude that such
> > > significant numerical errors cannot be justified even from a qualitative
> > > point of view.  Would you?
> > >
> > > Researchers in non-linear dynamical systems have recognized that it is
> > > difficult to compute chaotic solutions of non-linear differential equations
> > > for almost ten years.  For a dissipative system, there are attractors,
> > > but they have found that the properties of attractors due to different
> > > difference schemes are not the same.  As a result, they suggested that one
> > > should redo the computation with different time steps and compare the rates
> > > of attraction if an attractor is observed in a computation.  If these rates
> > > vary for different time steps then the observed attractor is likely to be a
> > > numerical artifact. [IMA Journal of Numerical Analysis (2001), vol.21,
> > > 751-767].   This is the reason why these numerical errors are called
> > > "computational chaos," which has nothing to do with dynamical systems.
> > >
> > > An agreement with experimental data is insufficient to guarantee that a
> > > theory is correct.  We all know that there are more disagreements between
> > > numerical solutions and experiments in fluid mechanics.  This is why CFD
> > > is jokingly related to  as "Color Fluid Dynamics!"  It is also well known that
> > > turbulence theories come and go.  When they come, they all have been
> > > verified with experimental observation.  I believe that it is time for
> > > "computational turbulence" to go.
> > >
> > > I have read the articles in Annual Review of  Fluid Mechanics and Nature
> > > about Benard convection that you cited in your email, but I did not find
> > > that they show their results converge for different time steps.  This is
> > > an important check, but seems completely ignored in CFD.  Using a different
> > > time step, their results may not agree with the experimental observations.

> > >
> > > Sincerely,
> > >
> > > Lun-Shin Yao
> > >
> > > -----Original Message-----
> > > From: Friedrich Busse [mailto:Friedrich.Busse@uni-bayreuth.de]
> > > Sent: Thursday, January 17, 2002 5:12 AM
> > > To: Lun-Shin Yao
> > > Cc: 'busse@uni-bayreuth.de'; 'T.J.Pedley@damtp.cam.ac.uk'
> > > Subject: Re: your mail
> > >
> > >
> > > Dear Professor Yao,
> > > Your paper has not been rejected solely based on the referee's report,
> > > but also because I think that it is more appropriate for a journal of
> > > numerical analysis than for the J. Fluid Mech. Since the paper deals
> > > with the numerical analysis of chaotic solution of partial differential
> > > equations it has per se little to do with fluid dynamics. Fluid
> > > dynamicists are well aware of the difficulties they face in simulating
> > > turbulent flows. Fortunately the chaotic attractors with which they deal
> > > are quite robust such that simulations can describe the physics
> > > correctly even after long times inspite of failing classical criteria for
> > > numerical convergence. For shorter times numerical simulations can follow quite
> > > faithfully the chaotic solution as you can see from fig. 29 of the
> > > article of Bodenschatz et al. in Ann. Rev. Fluid Mech. 32:709-778, 2000, where
> > > experimental observations are compared with numerical simulations. See
> > > also Egolf et al. Nature 404:733-736, 2000.
> > > Sincerely yours, F. Busse
> > >
> > > On Tue, 15 Jan 2002, Lun-Shin Yao wrote:
> > >
> > > > Dear Professor Busse:
> > > >
> > > > I would like to thank you for the review of my paper to JFM, "non-linear
> > > > energy transfer: represented by the non-linear terms of partial differential
> > > > equations".
> > > >
> > > > I do not think the referee has read the paper carefully enough.  My
> > > > computation was done with 32 terms, and was not 4 or 5 terms as he
> > > > stated in his report.  What I presented is the results of a simple
> > > >  to computation show
> > > > that the numerical chaotic solution of a non-linear differential equation
> > > > does not converge.  We have also found this is true for the Lorenz equations
> > > > and Rossler equations.  It also has been shown by the Yorke group at
> > > > Maryland University that it is impossible to "shadow" a non-hyperbolic map

> > > > [Phys. Rev. Lett. 79, 59 (1997)].  I think it is time for the fluid dynamics
> > > > community to recognize that the "direct numerical solutions" for turbulence
> > > > are bounded errors and not the solutions of the Navier-Stokes equations.
> > > >
> > > > Sincerely,
> > > >
> > > > Lun-Shin Yao
> > > > Professor
> > > > Department of Mechanical & Aerospace Engineering
> > > > Arizona State University
> > > > Tempe, AZ 85287
> > > > Voice: (480) 965-5914
> > > >
> > >
> >
>